\documentclass[10pt,a4paper,aps,twocolumn,showpacs,superscriptaddress,nofootinbib]{revtex4-1}
\usepackage[colorlinks,hyperindex]{hyperref}
\usepackage{epsfig}
\usepackage{float}
\usepackage{float,upgreek,yfonts}
\RequirePackage{amssymb,amsmath}
\usepackage[latin1]{inputenc}
\usepackage{graphicx}
\usepackage{indentfirst}
\usepackage{color}
\begin{document}
\title{The nuclear configurational entropy approach to dynamical QCD effects}
\author{G. Karapetyan}
\email{gayane.karapetyan@ufabc.edu.br}
\affiliation{Centro de Ci\^encias Naturais e Humanas, Universidade Federal do ABC - UFABC, 09210-580, Santo Andr\'e, Brazil}
\begin{abstract}
This paper scrutinizes the dynamical QCD effects influence on mesons, namely, the mean-square root radius of a pion in the holographic light-front wave function setup, in the context of the AdS/QCD. The nuclear configurational entropy, associated to mesonic holographic light-front wave functions, is shown to have  a critical point that optimizes the two parameters of the spin-improved light-front wave function. The mean-square root pion radius and its cross-section, computed upon these derived values,  match the exact existing experimental data to the precision of 0.14\%, below the experimental error at the PDG.
\end{abstract}
\maketitle
\section{Introduction}

The configurational entropy (CE) is an important apparatus in studies  that involve
 the lattice quantum   chromodynamics (QCD) setup as well as the AdS/QCD approach that emulates high energy states in nuclear systems. The CE critical points reckon the dominance and the abundance of complex nuclear configurations.
Using the CE in the context of dynamical AdS/QCD models,  the production of high spin light-flavour mesons has been quantitatively obtained in Refs.  \cite{Bernardini:2016hvx,Barbosa-Cendejas:2018mng}, whereas Ref. \cite{Bernardini:2016qit} scrutinized the dominance of scalar glueball states with lower spin. Besides, critical density stellar distributions, modelled by Bose--Einstein condensates of gravitons, have been estimated,  using critical points of the CE
 \cite{Casadio:2016aum}.
Other interesting phenomena was investigated in the AdS/QCD holographic model, in Ref. \cite{Braga:2017fsb}, where the predominance of bottomonium  and charmonium  states was shown to have a more compressed underlying information, in the Shannon's theory sense. Moreover, the concept of  CE allowed to find out other phenomena, as the Hawking--Page transition  \cite{Braga:2016wzx}. Critical points of the CE also permitted to derive Korteweg--de Vries solutions in a (cold) quark-gluon plasma \cite{daSilva:2017jay}.
It must be mentioned that for all localized nuclear systems, it is possible to use the concept of CE in order to derive the crucial data, which can be used in the analysis of complex nuclear systems and characterize the behavior of strong interaction nuclear configurations in the framework of QCD theory \cite{Ma:2018wtw,Ma:2015lpa} and particle physics \cite{Alves:2017ljt,Alves:2014ksa}.

In general, the CE is determined by the energy density of a system, originally proposed in Refs. \cite{Gleiser:2012tu,Sowinski:2015cfa,Gleiser:2014ipa,Gleiser:2011di,Gleiser:2015rwa,Gleiser:2013mga}, which is spatially localized.
Recently, it has been shown in Refs. \cite{Karapetyan:2017edu,Karapetyan:2016fai} that, in the case of high-energy nuclear reactions, the Fourier transform of the  reaction cross sections, which characterize the probability that a nuclear reaction occurs, can be used instead of the energy density as the spatially localized function. In this case, the computed quantity was called the nuclear configurational entropy.

The CE is nowadays widely used in the field of QCD,  investigating the strong interaction
among particles and can describe most yields at high energy accelerators, representing an excellent agreement between experiment and phenomenology.
Dynamical spin effects in the holographic light-front wave function
of the pion, in the Color Glass Condensate (CGC) setup, was used to predict the mean pion radius, as well as the spacelike electromagnetic
form factor and the twist-2 pion distribution amplitude,  during the fitting iteration, in full agreement with the latest HERA data \cite{Ahmady-4}.
A quantum state, in a complete description, is represented by the set of wave functions which, in general, can depend on the spin,  momentum,  polarization, and color.
The light-front wave functions that outline an  effective nuclear system, satisfy a Schr\"odinger-like equation, whose eigenvalue is the  particle square mass. The light-front wave functions interconnect gluons and quarks to their  hadronic states, playing a key role in the hadronic scattering formalism. Particularly, it has prominent relevance on the description of the pion-nucleus diffractive scattering process, precisely reproducing the results of the LHCb experiment. The  distribution amplitudes and the  form factors that evaluate transitions  can also be evaluated by mesonic light-front wave functions. Using the pion spin-improved light-front wave functions, we shall
derive the root mean-square pion radius, studying the critical
points of the nuclear CE, constructed upon the spin-improved holographic wave function.

The present paper is divided as follows: Sect. II is devoted to describe the details of light-front wave functions phenomenology, which can be used to derive the main properties of hadrons in the framework of the AdS/QCD approximation. Sect. III gives an overview of the dynamical effects of light-front wave functions, representing the spin-improved expression for the holographic wave function, considering the main expression to determine the critical point of the nuclear system, using the nuclear CE  approach. It provides a reliable computation of the root mean-square pion radius, precisely matching experimental data \cite{pdg}. The conclusion and main perspectives are presented in Sect. IV.

\section{The hadron light-front wave function}

The hadron light-front wave function can be derived from the light-front  Heisenberg equation for QCD  \cite{Ahmady-4}, 
\begin{equation}
H |\Psi\rangle = M^2 |\Psi \rangle,
\label{1op}
\end{equation}
where $H$  is the (light-front QCD) Hamiltonian, $M$ denotes the hadron mass associated to the $|\Psi \rangle$ hadronic state.
Given the spacetime metric,  $ds^2 = c^2dt^2 - d{\rm x}^2 - dy^2 - dz^2$, light-cone coordinates
are introduced, ${\rm x}^\pm = \frac{1}{\sqrt{2}}(ct \pm x)$, leading to the metric $ds^{2}=2d{\rm x}^{+}d{\rm x}^{-}+\delta _{ij}d{\rm x}^{i}d{\rm x}^{j}$, analogously for other quantities.

The hadron wave function depends on the transverse momentum $\mathbf{k}_{\perp i}$,  the individual momentum $k_j$ of the constituents, the fraction of momenta $x_j=k_j^+/P^+$ as well as the helicity $h_j$.
The hadron is represented in the Fock space at equal light-front time $({\rm x}^+=0)$ by
\begin{widetext}
\begin{equation}
|\Psi \rangle = \sum_{n,h_j} \int {d} {\rm x}_j \, {d}^2 {\mathfrak{k}}_{\perp j}\,x_j^{-1/2}\Psi_{n}(x_j,\mathbf{k}_{\perp j},h_j) \;|\;n: x_j P^+, x_j \mathbf{P_{\perp}} + \mathbf{k}_{\perp j}, h_j \rangle,
\label{1234}
\end{equation}
\end{widetext}
where the measures \begin{equation}
\!\!\!{{d} {\rm x}_j}\!=\!\prod_{i}^{n} \!{d} x_j \delta \!\!\left(\!1\!-\!\!\sum_{j=1}^{n} x_j\!\!\right),\; {d}^2 \mathfrak{k}_{i}\!=\! \prod_{i=1}^{n}\! {d}^2 \mathbf{k}_j\delta^2\!\!\left(\sum_{j=1}^n \mathbf{k}_j\!\right).
\label{3op}
\end{equation}
take into account the averages over selected constituents  \cite{Ahmady-4}. The hadronic wave function is also dependent on $\mathbf{P_{\perp}}, P^+, S_z$,
where $\Psi_{n}(x_j, \mathbf{k}_{\perp i},h_j)$ is the Fock space wave function with $n$ constituents. The semiclassical approach suggests that the quarks masses are disregarded, being  the hadronic wave function solely dependent  on the invariant mass $\mathcal{M}^2=(\sum_{i}^n k_j)^2$.

Denoting by $q$ a quark and $
\mathfrak{X}(x)=\sqrt{x(1-x)},$
when mesonic states are regarded, the  meson $q\bar{q}$ has mass $\mathcal{M}_{q\bar{q}}^2=k_{\perp}^2/\mathfrak{X}^2(x)$ \cite{Ahmady-4}. The term $\upzeta^2=\mathfrak{X}^2(x) b^2$ denotes the Fourier conjugate of the impact variable with parameter $b$, measuring the quark-antiquark separation.
Then, the valence light-front wave function is given by:
\begin{equation}
\Psi(\upzeta, x, \phi)= \exp({iL\phi})\frac{\phi (\upzeta)}{\sqrt{2 \pi \upzeta}}  \mathfrak{X}(x)
\label{4op}
\end{equation}
Eq. \eqref{1op} becomes an 1D Schr\"odinger-like equation, for the propagation of the wave functions of spin-$J$ for the transverse mode, wherein the confinement potential $U(\upzeta)$ consists of the set of all interaction terms that contain higher Fock states.
In the context of the AdS/QCD setup, in the bulk anti-de Sitter space (AdS) space, the confinement potential is constructed upon the impact light-front variable $\upzeta$, identified with fifth dimension of AdS, being nothing more than an energy scale. Besides,
the orbital angular momentum reads $L^2 = m^2R^2-(J-2)^2$, with $R$ and $m$ respectively being the AdS radius and mass, correspondingly. It leads to:
\begin{equation}
U(z_5, J)\!=\upphi''(z_5) \!+\! \frac{1}{4}\upphi'(z_5)^2 + \frac{2J-3}{4 z_5} \upphi'(z_5),
\end{equation}
where $\upphi(z_5)$ denotes the dilaton field \cite{csaki}.  In the dynamical AdS/QCD, the dilaton is quadratic on the energy scale, $\upphi(z_5)=\upkappa^2 z_5^2$, where $\upkappa$ is the the bulk coupling constant  \cite{Bernardini:2016hvx}. Hence, the confining potential reads $
U(\upzeta,J)= \upkappa^2 (\upkappa^2\upzeta^2 + J-1).$
Solving the Schr\"odinger equation, one can get the meson mass spectrum \cite{csaki} $
M^2= 4\upkappa^2 \left(n+\frac{L+J}{2}\right)$. Therefore the Schr\"odinger-like equation eigenfunctions are Laguerre polynomials,
 \begin{equation}
\phi_{n,L}(\upzeta)= \upkappa^{L+1} a_{n,L} \upzeta^{L+\frac12} e^{{{-\upkappa^2 \upzeta^2}/{2}}} L_n^L(\upzeta^2x^2 ) \;,
\label{laguerr}
\end{equation}
where $a_{n,L} =\sqrt{\frac{2 n !}{(n+L)!}}$.
One can entirely estimate the holographic wave function given by Eq. (\ref{laguerr}) by determining the form of longitudinal mode, \begin{widetext}
\begin{equation}
\Psi_{n,L}(\upzeta, x, \phi)= e^{iL\phi} \sqrt{\mathfrak{X}(x)} (2\pi)^{-1/2}\upkappa^{1+L} a_{n,L} \upzeta^{L} e^{{-\upkappa^2 \upzeta^2}/{2}} L_n^L(x^2 \upzeta^2).
\end{equation}
\end{widetext}
In the case of a pion with accounting of the quark mass, Eq. \eqref{4op} leads to:
\begin{equation}
\Psi^{\uppi} (x,\upzeta) = \mathcal{N} \sqrt{\mathfrak{X}(x)}  e^{ -{ \upkappa^2 \upzeta^2/ 2} }
e^{-{m_f^2/( 2 \upkappa^2\mathfrak{X}(x)) }}
\label{6opp}
\end{equation}
with normalization constant $\mathcal{N}$ computed by the probability $P_{q\bar{q}}$  for the meson to be described by a quark-antiquark Fock state, as $P_{q\bar{q}}=
\int {d} x\, {d}^2 \mathbf{b}\,  |\Psi^{\uppi}(x,\upzeta)|^2.$
As it was demonstrated in Ref. \cite{Ahmady-4}, it is possible to describe the pion observables by a universal $\upkappa$ AdS/QCD scale, without involving higher Fock states. It means that the momentum-dependent helicity wave function,   taking into account dynamical spin effects, must be used. This energy scale  can be chosen to fit the experiments, being $\upkappa=590$ MeV [523 MeV] for pseudoscalar [scalar] mesons \cite{bro}.

\section{Dynamical spin effects, the pion radius and nuclear configurational entropy}

In order to include the helicity dependence into  the holographic wave function,  the wave function $\Psi(x,\mathbf{k})$ must be corrected as $\Psi_{h\bar{h}}(x, \mathbf{k}) = S_{h\bar{h}}(x, \mathbf{k})  \Psi(x,\mathbf{k})$, for $S_{h\bar{h}}(x,\mathbf{k})$ being the helicity wave function for a $q\bar{q}$ meson.
Mesonic states present a form analogous to the meson-photon coupling, yielding
\begin{eqnarray}
S_{h\bar{h}}^{V}(x, \mathbf{k})=\frac{\bar{{\rm v}}_{\bar{h}}}{\sqrt{(1-x)}} \gamma_\mu \epsilon^\mu_V \frac{{\rm u}_{h}}{\sqrt{x}}.
\label{10opp}
\end{eqnarray}
where $
\bar{{\rm v}}_{\bar{h}}=
\bar{v}_{\bar{h}}((1-x)P^+\!,\! -\mathbf{k})
$ and ${\rm u}_{h}=u_{h}(xP^+\!, \mathbf{k}){\sqrt{x}}$ are components of the standard Dirac spinor, whereas the $\gamma_\mu$ are the Dirac matrices. The term $\epsilon_V^{\mu}$ denotes the mesonic  polarization vector.
It should be mentioned that, in the case of a pseudoscalar pion, the product $\gamma_\mu \epsilon^\mu_V$
in Eq. \eqref{10opp} must be replaced by a scalar field, that is a function of the pion momentum, $A (P^\mu \gamma_\mu)  + B \sqrt{P^\mu P_\mu}$, for $A$ and $B$ scalars, multiplied by the $\gamma^5$ chiral operator.  Therefore, the helicity wave function reads
\begin{eqnarray}
S^{\uppi}_{h\bar{h}}(x, \mathbf{k}) &=&\frac{{\rm \bar{v}}_{\bar{h}}}{\sqrt{1-x}} \left[(A \gamma^\mu P_\mu + B M_{\uppi}) \gamma^5 \right] \frac{{\rm u}_{h}}{\sqrt{x}}\label{110opp}\\
\label{spinwf}
\!\!\!\!\!\!\!\!\!\!\!\!\!\!\!\!\!\!\!\!\!\!\!\!\!\!\!\!\!\!\!\!&\!\!\!\!\!\!\!\!\!\!\!\!\!\!\!\!\!\!\!\!\!\!=&\!\!\!\!\!\!\!\!\!\!\!\!2h \left(\!A M_{\uppi}^2 \!+\! B \frac{m_f M_{\uppi}}{\mathfrak{X}^2(x) }\!\right)\! \delta_{-h \bar{h}} \!+\! B \frac{M_{\uppi} k e^{2hi \theta_k}}{\mathfrak{X}^2(x)} \delta_{h \bar{h}} ,\nonumber
\label{110opp}
\end{eqnarray}
using the spin sums rules in Ref. \cite{Lepage},
where $\mathbf{k}=k e^{i\theta_{k}}$.
The helicity wave function, being spin-dependent, assumes the existence of configurations with the both the same and the opposite helicity for the $q\bar{q}$ coupling.
The dynamical spin effects therefore require  some alteration due to the spin-orbit term in the Schr\"odinger-like equation.
Therefore, the Fourier transform of the spin-dependent holographic light-front wave function implies that
\begin{widetext}
\begin{equation}
\Psi_{\uppi}^{h\bar{h}}(x, \mathbf{b}) =  \left[ 2h M_\uppi(A \mathfrak{X}^2(x) M_{\uppi}  + B m_f ) \delta_{-h \bar{h}} -  iB M_{\uppi} \delta_{h \bar{h}}\partial_b   \right]\frac{\Psi_{\uppi} (x,\upzeta)}{\mathfrak{X}^2(x)} .
\label{sspin}
\end{equation}
\end{widetext}
When comparing Eq. \eqref{sspin} to the original holographic wave function, $
\Psi_{\uppi}^{h\bar{h}}(x,\mathbf{b})=  \frac{1}{\sqrt{2}}  h \delta_{-h \bar{h}} \Psi_{\uppi}(x,\upzeta)$, one can realize the component  $\Psi_{\uppi}(x,\upzeta)$  is given by Eq. \eqref{6opp}, whose normalization $\mathcal{N}$  can be computed by requiring that
\begin{equation}
\!\int \!\!{d} x\, {d}^2 \mathbf{b} \, |\Psi^{\uppi} (x,\mathbf{b})|^2 \!=\! 1,\; \text{for $
|\Psi^{\uppi}(x,\mathbf{b})|^2\!=\! \sum_{h,\bar{h}}
|\Psi^{\uppi}_{h\bar{h}} (x,\mathbf{b})|^2$}.
\end{equation}

Relevant features of a meson, as its radius and cross-section, can be estimated through the (spin-improved) holographic wave function. A non-perturbative approach, thus, yields the pion radius using the light-front wave functions, for the quark masses catalogued in the PDG \cite{pdg}.
The pion root mean-square radius can be estimated as \cite{Ahmady-4}:
\begin{equation}
\sqrt{\langle r_{\uppi}^2 \rangle} = \left[\frac{3}{2} \int {d} x\, {d}^2 \mathbf{b}\,|\Psi^{\uppi}(x,\mathbf{b})|^2\, b^2 (1-x)^2  \right]^{1/2}.
\label{radius}
\end{equation}
This expression was emulated by the  measured experimental values \cite{pdg}, using the original holographic wave function \eqref{6opp} and the spin-improved one \eqref{sspin}.

Now, the associated pion cross section
can be considered to construct and compute the nuclear CE \cite{Karapetyan:2018oye,Karapetyan:2017edu,Karapetyan:2016fai}, using the spin-improved  holographic light-front wave function. There are numerical and computational studies
which are devoted to investigation of various parameters obtained during experimental and phenomenological routine and which were predicted by using the CE  techniques in the context of the AdS/QCD. Although the dominance of lower spin glueballs was corroborated by the CE, whereas quarkonia $S$-wave states \cite{Braga:2017fsb} and radiation \cite{Braga:2016wzx} were also studied by the CE, and the abundance of lower spin light-flavour mesonic states was derived in \cite{Bernardini:2016hvx,Barbosa-Cendejas:2018mng}, no prediction has been derived out of the CE, regarding the inner mesonic structure. This is our main aim in what follows.

The first thing that must be accomplished, in order to establish the nuclear CE scheme,
is to calculate the spatial Fourier transform of the cross section of the pion, based on the spin- dependent holographic light-front wave function (\ref{sspin}).
In the original concept of the CE \cite{Gleiser:2012tu,Gleiser:2014ipa,Gleiser:2011di,Gleiser:2013mga}, one uses the energy density, which is spatially-localized observable, for computing the CE. However, in many cases, mainly the ones concerning nuclear systems, it is more convenient to employ the  square-integrable  cross section, which is also spatially-localized, in order to define the nuclear states. Therefore  the formula for the nuclear CE can be obtained \cite{Karapetyan:2018oye,Karapetyan:2017edu,Karapetyan:2016fai}, by first calculating the Fourier transform of the nuclear cross section,
\begin{equation}
\label{34}
{\upsigma}{(k)}=\frac{1}{2\pi}\! \int_{-\infty}^\infty\upsigma (b)\, e^{ikb} d b\,\,,
\end{equation}  with the respective modal fraction,
\begin{equation}\label{modall}
{\rm f}_{\upsigma(k)}=\frac{\vert\,\upsigma(k)\,\vert^2}{\int_{-\infty}^\infty\vert\,{\upsigma(k)\vert\,^2}\,dk}.
\end{equation}
 Finally, the nuclear configurational entropy is calculated by  \cite{Gleiser:2012tu}:
\begin{equation}
\label{333}
S_c \,= \, - \int_{-\infty}^\infty {\rm f}_{\upsigma(k)} \ln  {\rm f}_{\upsigma(k)} d k.
\end{equation}

Now Eqs. (\ref{34} - \ref{333}) for the nuclear CE can be computed for the spin-improved pion holographic wave function \eqref{sspin}.
Then Eqs. (\ref{34} - \ref{333}) are employed, in order to calculate the pion root mean-square radius by Eq. \eqref{radius}.
For the AdS/QCD energy scale
$\upkappa=523$ MeV \cite{Ahmady-4} and the  constituent quark masses at the PDG \cite{pdg}, the sets
a) $A=0,B=1$; and b) $A=1,B=1$, were adopted for the spin-improved wave function in Ref. \cite{Ahmady-4}.
However, we here adopt $A,B$ in Eq. (\ref{110opp}) as the two free parameters that shall be determined by the critical points of the nuclear CE, a posteriori.
The results obtained for the nuclear CE, for the holographic
spin-improved wave function, show an excellent agreement with experimental data \cite{pdg}, further corroborating to the estimation achieved with the assumption $A=1=B$ in Ref. \cite{Ahmady-4}.
Numerically calculated by Eqs. (\ref{34} - \ref{333}), using Eq. (\ref{333}), the nuclear CE  is then plot in Fig. 1.

\begin{figure}[!htb]
       \centering
                \includegraphics[width=2.5in]{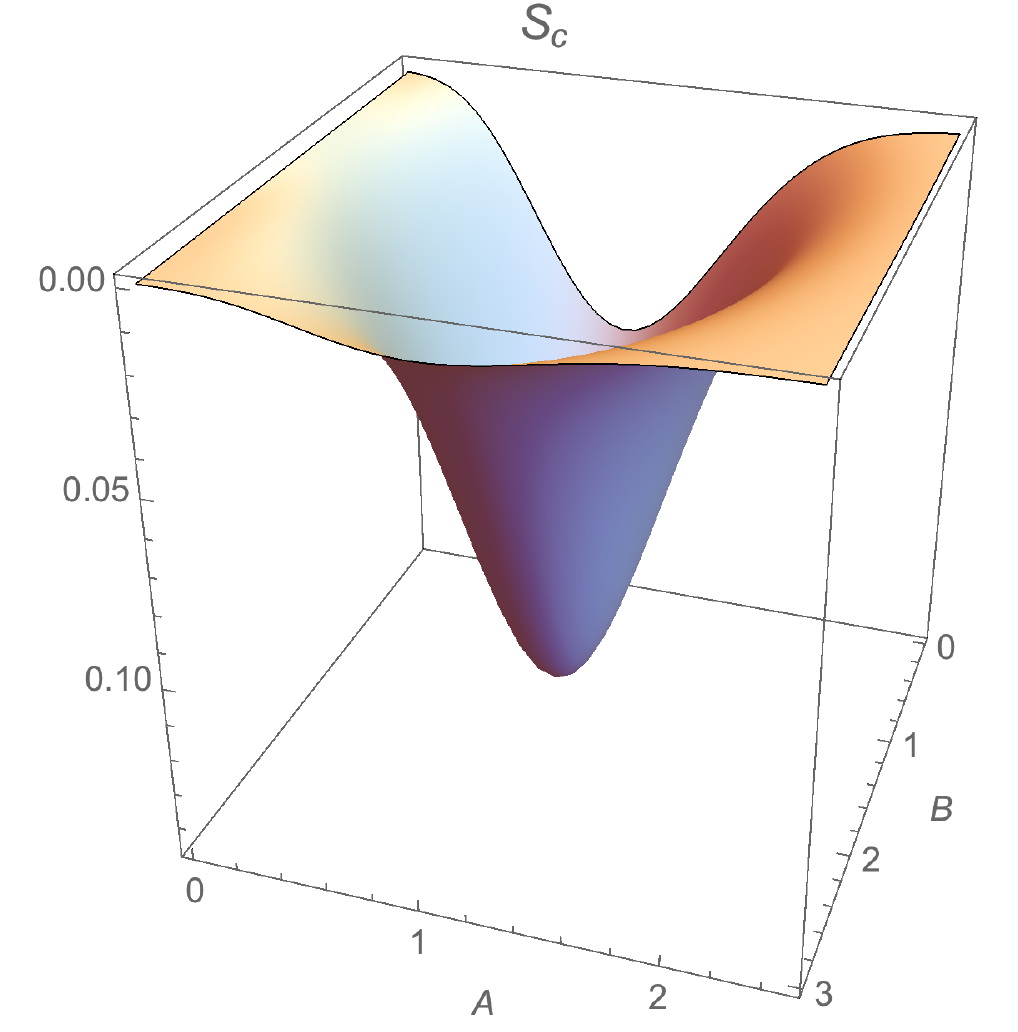}
                \caption{\textcolor{black}{Nuclear configurational entropy $S_c$ as a function
                of the $A,B$ parameters that appear in the spin-improved holographic wave function (Eq. (\ref{110opp}))}.}
                \label{mesons1}
\end{figure}
The critical point of the nuclear CE in Fig. 1 corresponds to $A=0.989$ and $B=1.010$,
showing a deviation of at most 1.1\% from the assumption $A=1=B$ in Ref. \cite{Ahmady-4}.
Now, going back to the spin-improved holographic wave function in  Eq. (\ref{110opp}) with $A=0.989$ and $B=1.010$, we can compute the spin-improved pion holographic wave function. Subsequently,  the pion root mean-square radius is numerically computed by Eq. (\ref{radius}), yielding
\begin{equation}
\sqrt{\langle r_{\uppi}^2 \rangle} = 0.673\, {\rm fm},
\label{radius}
\end{equation}
This result has full compliance to the experimental PDG data for $\sqrt{\langle r_{\uppi}^2 \rangle} = 0.672\pm 0.008\, {\rm fm}$  \cite{pdg}, within a 0.14\% precision, lower than the experimental error \cite{pdg}. 

Using the concept of the Shannon's information entropy \cite{Gleiser:2011di}, one can figure out the critical point of the nuclear CE and thus establish the natural selection of the observable pion root mean-square radius. This procedure corroborates
with the theoretical assumption $A=1=B$ in Ref. \cite{Ahmady-4} and with the experimental data in  PDG \cite{pdg}. Such set of variables, $A=0.989$ and $B=1.010$, were determined by the critical point of the nuclear CE, providing
a platform in nuclear and quantum system for the predicted value of the pion radius.
The critical point of the nuclear CE correspond to the dominant states of the nuclear configurations.

\section{Conclusions}

In the framework of the AdS/QCD, the parameters of the pion radius habe been computed, using concept of the nuclear  CE. 
Eqs. (\ref{34} - \ref{333}) have been applied to compute the critical points of the nuclear CE, as a function of the two free parameters $A$ and $B$ in the spin-improved holographic wave function.
The obtained result $A=0.989$ and $B=1.010$  corroborates the employed set of parameters 
$A=1=B$ suggested in Ref. \cite{Ahmady-4} and based on the experimental data analyses for
the holographic spin-improved wave function.
Using the above mentioned parameters, one can get the spin-improved pion holographic wave
function to compute the
pion root mean-square radius.
The result of radius of pion $\sqrt{\langle r_{\uppi}^2 \rangle} = 0.673\;{\rm fm}$ \
is in an excellent agreement with experimental value 
$\sqrt{\langle r_{\uppi}^2 \rangle} = 0.672\pm 0.008\ {\rm fm}$ \cite{pdg}.
During the calculation procedure the critical point of the nuclear CE predicts the predominant nuclear states, providing the natural set of the observables and show an excellent agreement not only with theoretical and phenomenological predictions but also with experimental data.
Our calculations also corroborates to the results obtained by the procedure fitting in Ref. \cite{Ahmady-4}, where
simulations were accomplished using a spin-improved light-front
holographic wave function for the pion together with a
universal AdS/QCD scale.
The meson description by a quark-antiquark Fock state, coupled with the holographic light-front meson wave function, serves as an appropriate mode to predict the pion radius in the framework of the QCD theory.
Hence, one can conclude that the nuclear configurational entropy can be, once more, employed as a reliable tool to obtain the fundamental parameters of nuclear reactions. 
We intend to continue the study of nuclear configurations by involving, also, improved wave functions in the hadronic applications of quantum mechanics, based on the proposals in Refs. \cite{Bazeia:2013usa,Bazeia:2012qh,Bernardini:2016rgb,Bazeia:2012qh,Bernardini:2012bh}, in the CE framework \cite{Correa:2016pgr,Correa:2015vka}.

  \acknowledgements
GK thanks to FAPESP (grant No. 2016/18902-9), for partial financial support.

\end{document}